\documentclass[twoside,fleqn]{ActaStyle}

\usepackage{amsmath}
\usepackage{amssymb}
\usepackage{epsfig}

\newcommand{\pt}{p_T}

\def\authorheading{R.~Lietava et al.}

\def\shorttitle{Jet quenching in nuclear collisions}

\begin{document}

%%%%%%%%%%%%%%%%%%%%%%%%%%%%%%%%%%%%%%%%%%%%%%%%%%%%
%%%%%%%%%%%%%%%%%%%%%%%%%%%%%%%%%%%%%%%%%%%%%%%%%%%%
% preprint number, only for posting in the TH division
%\begin{flushright}CERN-TH/2003-210, nucl-th/0309014\end{flushright}
%%%%%%%%%%%%%%%%%%%%%%%%%%%%%%%%%%%%%%%%%%%%%%%%%%%%
%%%%%%%%%%%%%%%%%%%%%%%%%%%%%%%%%%%%%%%%%%%%%%%%%%%%

%%%  page range, first and last page
\pagerange{1}{11}%\pageref{lastpage}

%%%%%%%%%%%%%%%%%%%%%%%%%%%%%%%%%%%%%%%%%%%%%%%%%%%%%%
\title{%
A MODEL OF BINARY COLLISIONS DEPENDENCE OF JET QUENCHING
IN NUCLEAR COLLISIONS AT ULTRARELATIVISTIC ENERGIES}

\author{%
Roman Lietava$^*$, J\'an Pi\v s\'ut\email{Jan.Pisut@fmph.uniba.sk}$^\dagger$, 
Neva Pi\v s\'utov\'a$^\dagger$, Boris Tom\'a\v sik$^{\ddagger\diamondsuit}$}
{%
$^*$ School of Physics and Astronomy, University of Birmingham,\\ 
Birmingham B15 2TT, UK\\
$^\dagger$ College of Mathematics, Physics and Information Technology,\\ 
Comenius University, Mlynsk\'a Dolina, SK-84248 Bratislava, Slovakia\\
$^\ddagger$ CERN, Theory Division, CH-1211 Geneva 23, Switzerland\\
$^\diamondsuit$ The Niels Bohr Institute, Blegdamsvej 17, 
DK-2100 Copenhagen \O , Denmark
}

%%% Date of submission
\day{September 8, 2003}

\abstract{%
We describe a model of jet quenching in nuclear collisions 
at RHIC energies. In the model, jet quenching is to be caused
by the interruption of jet formation by nucleons arriving at the 
position of jet formation in a time shorter than the 
jet formation time. Our mechanism predicts suppression of
high-$\pt$ spectra also in d+Au reactions.
}

%%% PACS numbers
\pacs{%
25.75.-q, 25.45.-z, 25.75.Nq
}

%%%%%%%%%%%%%%%%%%%%%%%%%%%%%%%%%%%%%%%%%%%%%%%%%%%%%%%%%%%%%%%%%
%%%%%%%%%%%%%%%%%%%%%%%%%%%%%%%%%%%%%%%%%%%%%%%%%%%%%%%%%%%%%%%%%
\section{Introduction}
\label{intro}

Hadronic spectra at high transverse momenta from ultrarelativistic 
nuclear collisions result from jets produced 
in the early hard partonic interactions. The yields 
are normally expected to scale with the number of binary nucleon-nucleon
collisions. However, experimental results obtained at the Relativistic
Heavy Ion Collider (RHIC) contradict these expectations: in central Au+Au
collisions the yields above 2~GeV/$c$ are {\em strongly suppressed} 
with respect to binary collisions scaling 
\cite{starAuAu02,starAuAu03,phenixAuAu02,phenixAuAu03}.
Recent data from d+Au collisions show an {\em enhancement}
of the yield with respect to the binary-collision scaling for 
$\pt \gtrsim 2\, \mbox{GeV}/c$ \cite{phenixdAu,stardAu,phobosdAu,brahmsdAu} 
but indicate that the yield may fall below this expectation above 
8~GeV/$c$ \cite{stardAu}. 

The lack of strong suppression of high-$\pt$
yields in deuteron-induced reactions suggests that the effect in Au+Au
collisions is due to partonic energy loss in a medium with 
very high energy density. This mechanism has been studied by many authors
\cite{WGP,Wie,GLV,VGL,HN,Mul}. 

In addition, there are some models which aim to describe 
high-$\pt$ spectra from nuclear collisions by invoking
different physical effects. The perturbative-QCD-improved parton 
model of refs.~\cite{zhang,papp} 
includes shadowing and the broadening of $\pt$ spectra due to 
intrinsic parton transverse momentum. The authors conclude that 
the mechanism is insufficient to explain high-$\pt$ 
suppression at RHIC and even at the SPS.

The mechanism of parton saturation \cite{GLR,MQ,MV}
was argued to lead to suppression 
of intermediate $\pt$ production in both Au+Au and d+Au collision 
systems \cite{KLM,KLN}.

In ref.~\cite{LPPT1}, we proposed a model for the suppression of
high-$\pt$ spectra in ultrarelativistic nuclear collisions based 
on arguments involving {\em the uncertainty principle} and {\em formation time}
\cite{PP97,FP56}. The mean-free path of an incident nucleon 
in the nucleon--nucleon centre-of-mass system (CMS)  
at RHIC is rather short: $\lambda \approx 0.025$~fm. 
When two jets are
produced in a particular nucleon--nucleon collision, the next
nucleon arrives at the position where the jets were created in a time
interval of the order of $\lambda/c$. 
According to the 
uncertainty relation, a process with longitudinal momentum 
transfer $\Delta p_L$ and energy transfer $\Delta E$ needs
space and time intervals given by
\begin{equation}
\Delta z > \frac {\hbar}{\Delta p_L}\, , \qquad
\Delta t > \frac {\hbar}{\Delta E}\, .
\label{eq1}
\end{equation}
to reach completion.
Hence, if the mean-free path or the mean-free time are shorter than 
these intervals, the created jets may be seriously influenced 
by nucleons arriving at the position where the process develops.
For the quoted mean-free path at RHIC, this puts a limit on 
processes which are {\em not} influenced by our mechanism at 
$\Delta p_L \gtrsim 8\,\mbox{GeV}/c$. 
The limit is, of course, not strict, 
since the time interval between subsequent collisions fluctuates.

In the next Section we present a slightly improved version of the model 
from our earlier paper \cite{LPPT1}. The results are shown in 
Section~\ref{results} and we conclude in Section~\ref{concl}.
The appendix contains a discussion of Lorentz invariance in our model.
In what follows we shall work in natural units $c = \hbar = 1$.

%%%%%%%%%%%%%%%%%%%%%%%%%%%%%%%%%%%%%%%%%%%%%%%%%%%%%%%%%%%%%%
%%%%%%%%%%%%%%%%%%%%%%%%%%%%%%%%%%%%%%%%%%%%%%%%%%%%%%%%%%%%%%
\section{The model}
\label{model}
%%%%%%%%%%%%%%%%%%%%%%%%%%%%%%%%%%%%%%%%%%%%%%%%%%%%%%%%%%%%%%

We shall formulate our Glauber model in the nucleon--nucleon 
centre-of-mass system. 
For the sake of simplicity, we assume that the density distribution 
is uniform. Technically, the model is rather similar to those
for nuclear absorption of $J/\psi$ in heavy-ion collisions \cite{GHRev}.

\begin{figure}[t]
\begin{center}
   \epsfig{file=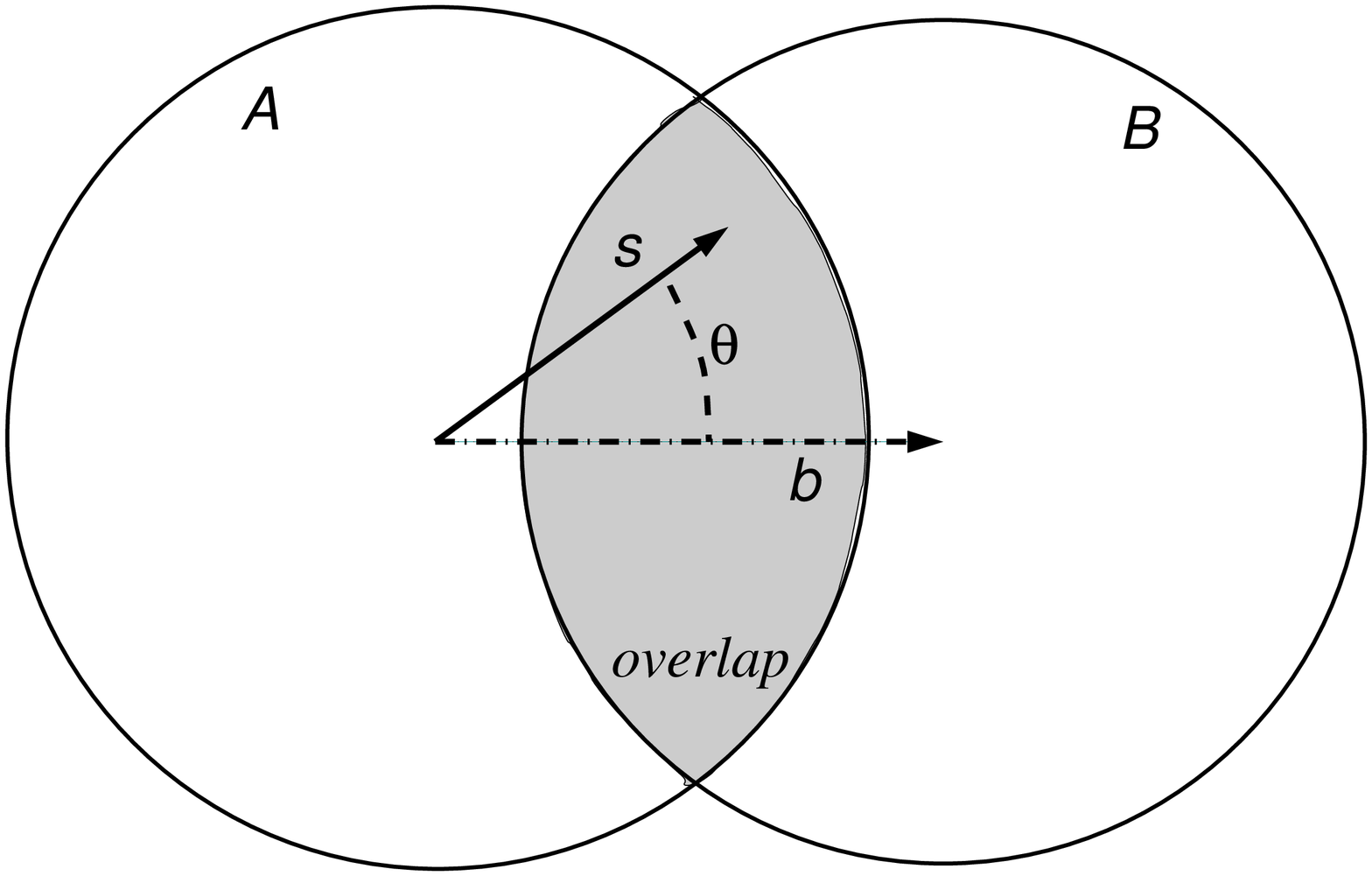,scale=0.26}\hspace{0.7cm}
   \epsfig{file=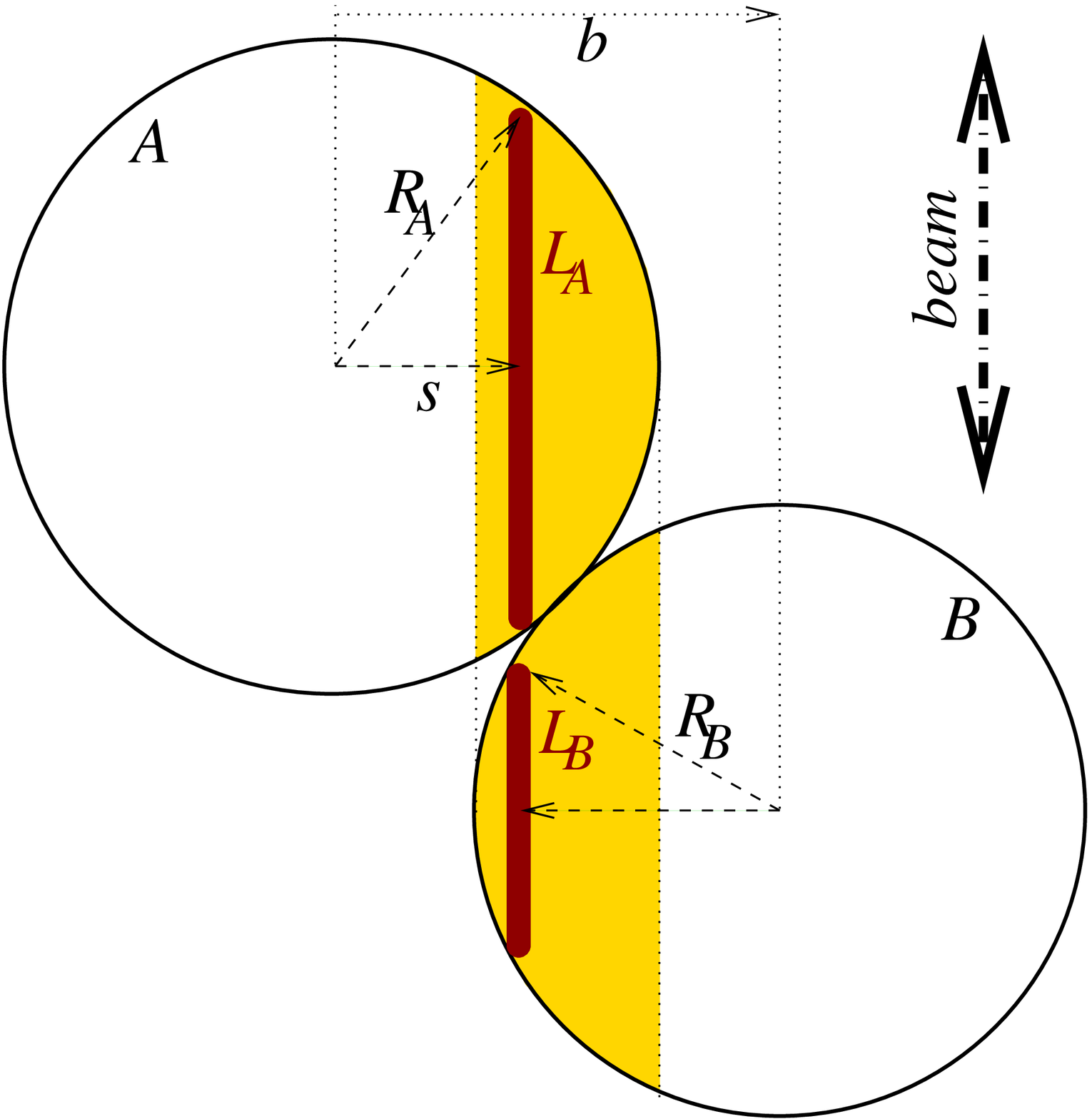,scale=0.3}
\end{center}
\caption{%
Left: geometry of non-central nuclear collisions. Right: 
layout of tube-on-tube interaction (plotted without Lorentz 
contraction).}
\label{F-layout}
\end{figure}

A Glauber model is usually formulated in terms of ``tube-on-tube''
collisions, with tubes filled with nucleons\footnote{The integral
of the density along the ``tube'', i.e.\ in the longitudinal direction,
is often called {\em the thickness function} 
$T_A(s) = \int_{-\infty}^{\infty} \rho(\sqrt{s^2 + z^2})\, dz$. In our model,
the density is given by  $\rho_A\theta(R_A-r)$ and
the thickness function corresponds to the tube length given in 
eq.~\eqref{tubes} multiplied by the constant nuclear density.}.
For a collision of nuclei A and B at impact parameter $b$ we have for the 
lengths of the colliding tubes
\begin{equation}
2\, L_A(s)=2\, \gamma^{-1} \sqrt{R_A^2-s^2}\, ,\quad 
2\, L_B(b,s,\theta)= 2\, \gamma^{-1}
\sqrt{R_B^2-b^2-s^2+2\, b\, s\cos\theta}\, .
\label{tubes}
\end{equation}
Here, $\gamma$ is the Lorentz contraction factor for the boost 
of the nuclei from their own rest-frame to the CMS. All other 
coordinates and sizes on the r.h.s., 
however, are taken in the rest frames of the 
nuclei and their meaning is explained in Figure~\ref{F-layout}.
%%%%%%%%%%%%%%%%%%%%%%%%%%%%%%%%%%%%%%%%%%%%%%%%%%%%%%%%%%%%%%%%%%%%%
%%%%%%%%%%%%%%%%%%%%%%%%%%%%%%%%%%%%%%%%%%%%%%%%%%%%%%%%%%%%%%%%%%%%%
The positions of nucleons within the colliding tubes will be denoted
by $z_A$ and $z_B$. The values of $z_A$ and $z_B$ satisfy
\[
-L_A \le z_A \le L_A, \quad -L_B \le z_B \le L_B.
\]
We take $z_A$ and $z_B$ as increasing in the direction
of motion of A and B respectively in the CMS.

We will be interested in comparing the yield of jets at high transverse
momentum produced in   nuclear collisions to that produced in  
nucleon--nucleon collisions. By staying at the level of jets and not 
including fragmentation, our calculation is not influenced 
by the form  of the fragmentation function which is not entirely 
known for nuclear collisions. On the other hand, we must keep in mind
that our results cannot be compared directly to the measured 
$\pt$-spectra.

The yield of jets in collisions of nuclei A+B at impact parameter $b$ 
is defined as
\begin{equation}
\label{yield_AB}
Y_{AB}(\pt,b) = 
\frac{ \frac{d\sigma_{AB}}{d\pt^2\, db^2}}{%
\frac{d\sigma_{AB}}{db^2}}\, ,
\end{equation}
while the yield in nucleon--nucleon collisions is introduced as
\begin{equation}
\label{yield_pp}
Y_{pp}(\pt) = \frac{ \frac{d\sigma_{pp}}{d\pt^2}}{\sigma_{pp}} \, .
\end{equation}

As in our previous paper \cite{LPPT1} we intend to determine
quantity
\begin{multline}
\label{Eq:R} 
R_{AB}(\pt,b) = \frac{Y_{AB}(\pt,b)}{Y_{pp}(\pt)} = \\
\sigma_{nn}
\int_{\rm overlap}s\, ds\, d\theta \, \int_{-L_A}^{L_A} dz_A \, \rho_A
\, \int_{-L_B}^{L_B} dz_B\, \rho_B   \, 
F(b,s,\theta,z_A,z_B)\, . 
\end{multline}
Here, $\rho_A$ and $\rho_B$ are the Lorentz-contracted nuclear densities
of the nuclei A and B, 
\[
\rho_A = \rho_B = \rho = \gamma\, \rho_0
\]
with $\rho_0 = 0.138\, \mbox{fm}^{-3}$. For the total inelastic 
nucleon--nucleon cross-section we will take 
\[
\sigma_{nn} = 40\, \mbox{mb} = 4\, \mbox{fm}^2\, .
\]
Without the suppression factor $F(b,s,\theta,z_A,z_B)$, formula
\eqref{Eq:R}
would give the average number of binary collisions
in an interaction of nuclei A+B at impact parameter $b$. The suppression 
factor will account for the assumed effect of jet destruction.

If jets are created in a hard interaction of two incident partons 
of the colliding nuclei, we shall assume a cross-section for their 
destruction by a subsequent incoming nucleon (in CMS) an expression
\begin{equation}
\sigma_a(p_T,t,t')=\sigma_0 \left( 
\frac{1}{1+(\pt (t'-t))^2}\right)^2 \, ,
\label{sig_a}
\end{equation}
where $t$ is the time coordinate of the hard interaction
and $t'$ is the time\footnote{%
In our previous paper \cite{LPPT1} we used the longitudinal 
coordinates $z$ and $z'$ instead of the times in our argumentation. 
The advantage of the new formalism is that $\sigma_a$ can be written 
in a Lorentz invariant way, as described in the Appendix. 
At the energies studied,
however, the numerical difference between $(t'-t)$ and $(z'-z)$ is 
negligible.}
when the destroying nucleon arrives at
the position of the jet\footnote{%
For brevity, we talk about  a "destroying nucleon" although 
we rather mean those of its quarks with large enough momentum. The problem 
with soft sea quarks is that they can be barely localised as precisely
as we need for our argument, but we can assume that their 
energy is too small to destroy the jet.}. 
The jet transverse momentum 
$\pt$ is roughly equal to the energy $\Delta E$ involved in the process.
Therefore, the form \eqref{sig_a} is in line with our considerations 
about the jet destruction based on the uncertainty relation described in
the introduction. The absorption cross-section is of order of few milibarns
and were tuned by the parameter $\sigma_0$.
Although formula \eqref{sig_a} for the cross-section is
written in the CMS reference frame, we demonstrate in the Appendix 
how it can be written in an explicitly invariant way.
The improvement over our previous paper \cite{LPPT1} lies in 
the time-dependent prescription for the absorption cross-section. 
In \cite{LPPT1} we ignored the fact 
that nucleons with different distances from the origin of the jet
have different chances to destroy it, and we assumed 
\begin{equation}
\sigma_a^{\rm old}(\pt)=\sigma_0 \left( \frac{1}{1+(p_T/p_{T0})^2}\right)^2
\qquad\qquad
\mbox{(the model of \cite{LPPT1})}
\, ,
\label{sig_a-old}
\end{equation}
with $p_{T0} = 8\, \mbox{GeV}/c$ for collisions at RHIC.

The suppression factor $F(b,s,\theta,z_A,z_B)$ now reads
\begin{multline}
\label{Eq:F}
F(b,s,\theta,z_A,z_B) = 
\exp\left ( - \int_{-L_A/v}^{{z_A}/{v}}
\sigma_a\left (\pt,t,\frac{z_A}{v}\right )\, \rho_A \, v\, dt\right )\\
\times \exp\left ( -\int_{-L_B/v}^{z_B/v}
\sigma_a\left (\pt,t,\frac{z_B}{v}\right )\,\rho_B \, v\, dt\right )
\, ,
\end{multline}
where $z_A$ and $z_B$ determine the position of the hard interaction
within nuclei A and B, respectively, and  $v$ is the longitudinal velocity 
of the colliding nuclei. The $b$, $s$, and $\theta$-dependences
of $F$ are implicitly included in $L_A$ and $L_B$. 
When we insert the prescription
\eqref{sig_a} into equation \eqref{Eq:F}, the integrals can be 
performed analytically and lead to
\begin{multline}
\label{Eq:Fint}
F(b,s,\theta,z_A,z_B) = \\
\exp\left [ - \frac{\sigma_0\, v\, \rho_A}{2\, \pt} 
\left ( \frac{f_A}{1+f_A^2} + \arctan f_A\right ) -
\frac{\sigma_0\, v \, \rho_B}{2\, \pt} 
\left ( \frac{f_B}{1+f_B^2} + \arctan f_B\right ) \right ]
\, ,
\end{multline}
with
\begin{equation}
\label{Eq:f}
f_A = \frac{\pt}{v}(z_A + L_A) \, , \qquad 
f_B = \frac{\pt}{v}(z_B + L_B) \, .
\end{equation}

We can compare our results with our previous model of ref.~\cite{LPPT1}. 
The suppression factor there was obtained by inserting  formula
\eqref{sig_a-old} for the absorption cross-section. This leads to
\begin{eqnarray}
\nonumber
F^{\rm old}(b,s,\theta,z_A,z_B) & = & 
\exp\left ( -\int_{-L_A}^{z_A} \sigma_a^{\rm old}(\pt)\, \rho_A\, dz \right)
\exp\left ( -\int_{-L_B}^{z_B} \sigma_a^{\rm old}(\pt)\, \rho_B\, dz \right) \\
& = & 
\nonumber
\exp \left ( - \sigma_a^{\rm old}(\pt)\, \rho_A\, (z_A+L_A) 
- \sigma_a^{\rm old}(\pt) \, \rho_B\, (z_B+L_B) \right ) \, .\\ &&
\label{F-old}
\end{eqnarray}
%

%%%%%%%%%%%%%%%%%%%%%%%%%%%%%%%%%%%%%%%%%%%%%%%%%%%%%%%%%%%%%%%%%
%%%%%%%%%%%%%%%%%%%%%%%%%%%%%%%%%%%%%%%%%%%%%%%%%%%%%%%%%%%%%%%%%
\section{Results}
\label{results}
%%%%%%%%%%%%%%%%%%%%%%%%%%%%%%%%%%%%%%%%%%%%%%%%%%%%%%%%%%%%%%%%%

In our calculation we simulated a set of nuclear collisions within 
the Glauber model framework and for every nucleon--nucleon interaction 
we determined the suppression factor according to 
eq.~\eqref{Eq:Fint}. By adding the suppression factors from 
all nucleon--nucleon interactions we obtained the total yield of jets relative
to the yield from nucleon--nucleon interactions as defined in
eq.~\eqref{Eq:R}. The set of nuclear collisions under 
study was always chosen appropriately to sample given centrality 
requirements.

The absorption cross-section parameter $\sigma_0$ was tuned to
\[
\sigma_0 = 8\, \mbox{mb} = 0.8 \, \mbox{fm}^2\, .
\]

In the figures we always plot the relative yield of jets divided by the 
number of binary collisions. In case of no suppression this 
number approaches unity. As we have already mentioned, we cannot compare our 
results directly to experimental data because fragmentation is not included
in our calculation. Nevertheless, we make some remarks on the relation of our 
results to data in Section~\ref{concl}.

\begin{figure}[t]
\begin{center}
   \epsfig{file=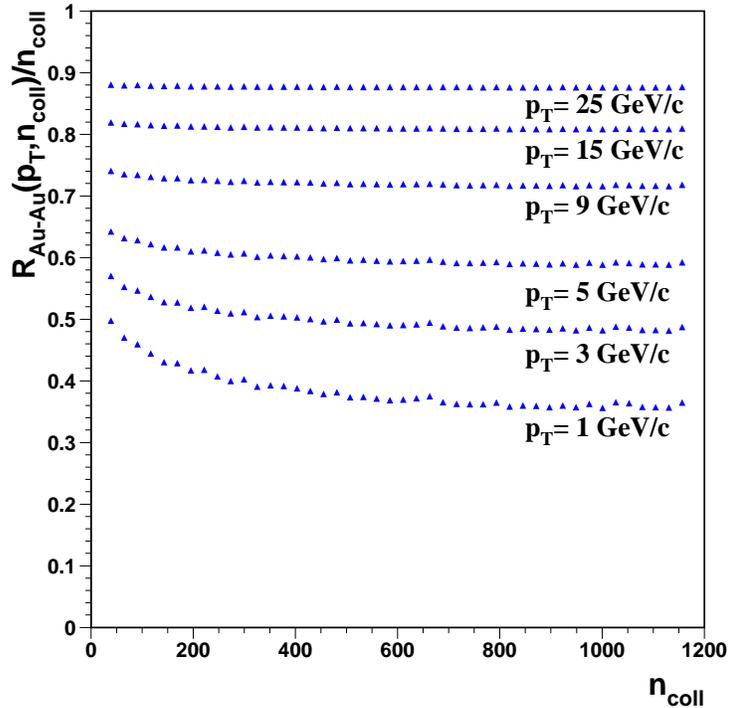,scale=0.5} 
\end{center}
\caption{%
The ratio $R_{\rm Au+Au}(\pt,b)/ n_{\rm coll}(b)$
for central collisions at $\sqrt{s} = 200~A\mbox{GeV}$
plotted as a function of $n_{\rm coll}$. Different curves correspond 
to variations of $\pt$, as indicated. 
}
\label{F:ncoll}
\end{figure}

Figure~\ref{F:ncoll} shows  the dependence of the relative yield
on the number of binary collisions.
%
%%%%%%%%%%%%%%%%%%%%%%%%%%%%%%%%%%%%%%%%%%%%%%%%%%%%%%%%%%%%%%%%%%%%%
%%%%%%%%%%%%%%%%%%%%%%%%%%%%%%%%%%%%%%%%%%%%%%%%%%%%%%%%%%%%%%%%%%%%%
%
We observe that scaling with the number of binary collisions 
is slowly recovered for large $\pt$. As the transverse momentum
increases, the curves converge to the asymptotic value of
$R_{AB}(\pt,n_{\rm coll})/n_{\rm coll} = 1$.

\begin{figure}[t]
\begin{center}
   \epsfig{file=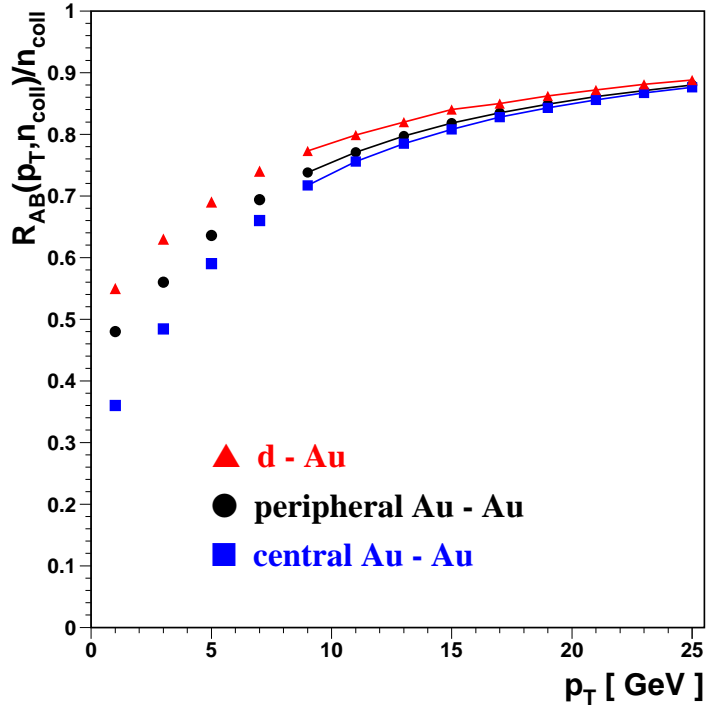,scale=0.5} 
\end{center}
\caption{%
The ratio $R_{\rm Au+Au}(\pt,b)/n_{\rm coll}(b)$
plotted as a function of the jet transverse momentum $\pt$. 
Calculated for collisions at $\sqrt{s} = 200~A\mbox{GeV}$.
Different curves refer to: 
central Au+Au collisions (0--5\% of the total cross-section),
peripheral Au+Au collisions (60--80\%), and minimum bias
d+Au collisions. 
}
\label{F:pt}
\end{figure}

The $\pt$-dependence of the yields of  jets is plotted in 
Figure~\ref{F:pt} for central and peripheral Au+Au collisions
and for d+Au collisions. 
%
%%%%%%%%%%%%%%%%%%%%%%%%%%%%%%%%%%%%%%%%%%%%%%%%%%%%%%%%%%%%%%%%%%%%%
%%%%%%%%%%%%%%%%%%%%%%%%%%%%%%%%%%%%%%%%%%%%%%%%%%%%%%%%%%%%%%%%%%%%%
%
Our mechanism leads to suppression of jet production even in 
d+Au collisions. We comment on the relation of this result to data
in the next Section.

\begin{figure}[t]
\begin{center}
   \epsfig{file=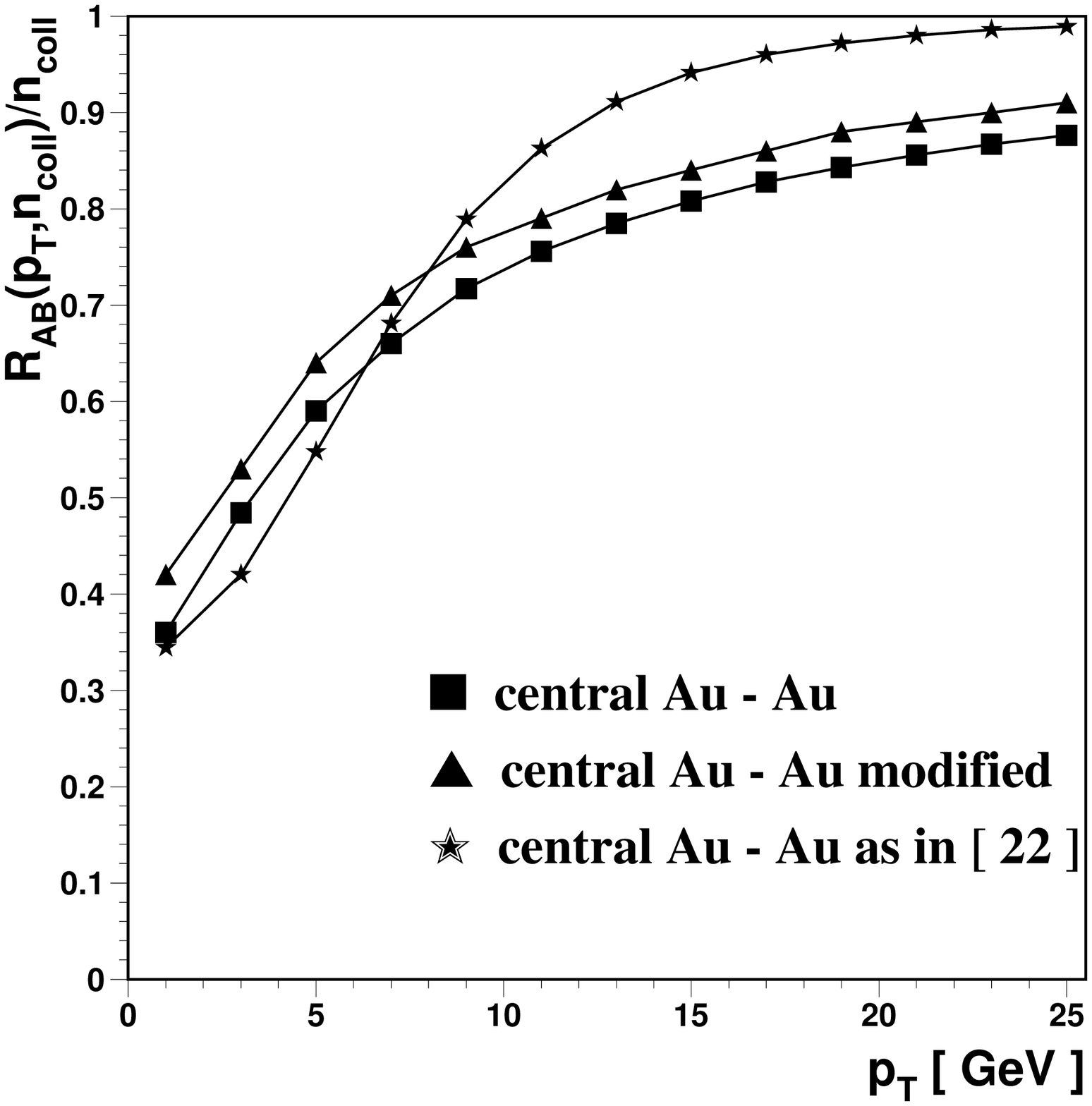,scale=0.5} 
\end{center}
\caption{%
The comparison of our model to that of our earlier paper \cite{LPPT1}.
The relative yield $R_{\rm Au+Au}(\pt,b)/n_{\rm coll}(b)$
plotted as a function of $\pt$, calculated for central
Au+Au collisions (0--5\% centrality cut) at $\sqrt{s} = 200~A\mbox{GeV}$.
}
\label{F:stare}
\end{figure}

In Figure~\ref{F:stare} we compare our model with that of ref.~\cite{LPPT1}.
%
%%%%%%%%%%%%%%%%%%%%%%%%%%%%%%%%%%%%%%%%%%%%%%%%%%%%%%%%%%%%%%%%%%%%%
%%%%%%%%%%%%%%%%%%%%%%%%%%%%%%%%%%%%%%%%%%%%%%%%%%%%%%%%%%%%%%%%%%%%%
%
This comparison shows that the models give similar results at low
transverse momenta while at high $\pt$ the suppression
is stronger in the new model.
This is due to destroying nucleons which may be very close to the jet 
production site in the new model (i.e., $(t'-t)$ is small)
and thus associated with a very large 
absorption cross-section. In the old model \cite{LPPT1}, 
on the other hand, all nucleons
were effectively put a distance $\lambda$ from the jet production 
such that there were effectively no nucleons that would be close enough 
to destroy very-high-$\pt$ jets%
\footnote{%
Mathematically, in the present model
the supression factor $F$ in eq.~\eqref{Eq:Fint} 
goes like $\exp(-\mbox{const}/p_T^2)$
for high $\pt$, while in the old model the factor $F^{\rm old}$ 
of eq.~\eqref{F-old} approaches $\exp(-\mbox{const}/\pt^4)$.}.
In Figure \ref{F:stare} we also show the result of a modified model
in which we forbid the nucleons within a nucleus (at rest) to be closer 
to each other
than 1~fm, such that there are no very close destroying nucleons.
This makes the suppression weaker by 5--15\%, depending on $\pt$.

When we compare the old and the actual model at low $\pt$,
we see the effect of suppression by subsequent nucleons 
folowing the first potentially destroying nucleon. While in the 
old model their absorption cross-sections were equal to that of the 
first destroying nucleon, in the new formulation these cross-sections
are smaller due to the larger distance from the  origin of the 
jet production. Therefore,
the new model gives  weaker suppression at low $\pt$ than the old one.

%%%%%%%%%%%%%%%%%%%%%%%%%%%%%%%%%%%%%%%%%%%%%%%%%%%%%%%%%%%%%%%%%
%%%%%%%%%%%%%%%%%%%%%%%%%%%%%%%%%%%%%%%%%%%%%%%%%%%%%%%%%%%%%%%%%
\section{Conclusions}
\label{concl}
%%%%%%%%%%%%%%%%%%%%%%%%%%%%%%%%%%%%%%%%%%%%%%%%%%%%%%%%%%%%%%%%%
We have described here an improved version of the jet attenuation model 
of \cite{LPPT1} in which the cross-section for the jet attenuation 
is time-dependent as implied by \eqref{eq1}. We have shown that 
results for the jet attenuation obtained by the new version 
do not differ very much from the previous version.

An important feature of our model is the lack of any
thresholds under which the mechanism of jet suppression 
would cease. We obtain a {\em suppression} with respect to the 
scaling with the number of binary collisions even in the case of 
{\em d+Au collisions}. This differs from 
the recently published high-$\pt$ data from d+Au
collisions at RHIC \cite{phenixdAu,stardAu,phobosdAu,brahmsdAu}.
An {\em enhancement}, i.e., a yield larger than expected 
from the $n_{\rm coll}$-scaling was reported from the 
deuteron-induced reactions. This is likely to be, at least in part, a 
Cronin effect which results from $\pt$-broadening of the incoming 
partons. Since there is no such effect included in the model, we 
cannot describe this feature.

However, the data on high-$\pt$ spectra of charged hadrons from
d+Au collisions 
published by STAR collaboration indicate \cite{stardAu} that 
the yield normalised to the $\pt$-spectrum and the number of binary
collisions  possibly falls below unity 
for $\pt \gtrsim 9\, \mbox{GeV}/c$. In this kinematic region, 
the Cronin effect may not be effective any more and the suppression 
could be described by our model. In order to indicate this,
in Figure~\ref{F:pt} we did not plot the curves  in the kinematic region 
where we expect the dominance of the Cronin effect and highlighted 
only the kinematic region where our model could stay relevant 
as is. We plan to combine the Cronin effect with the present model 
in the future and confront  our results with  existing data.

%%%%%%%%%%%%%%%%%%%%%%%%%%%%%%%%%%%%%%%%%%%%%%%%%%%%%%%%%%%%%%
%%%%%%%%%%%%%%%%%%%%%%%%%%%%%%%%%%%%%%%%%%%%%%%%%%%%%%%%%%%%%%
\section*{Acknowledgements}
%%%%%%%%%%%%%%%%%%%%%%%%%%%%%%%%%%%%%%%%%%%%%%%%%%%%%%%%%%%%%%

One of the authors (JP) is
indebted to the CERN Theory Division for the hospitality
extended to him. 
We like to thank Orlando Villalobos-Baillie
for many useful discussions and careful reading of the paper. 
The work of JP and NP was supported in part by the Slovak Ministry 
of Education under the Grant No.~VEGA V2F13.

%%%%%%%%%%%%%%%%%%%%%%%%%%%%%%%%%%%%%%%%%%%%%%%%%%%%%%%%%%%%%%
%%%%%%%%%%%%%%%%%%%%%%%%%%%%%%%%%%%%%%%%%%%%%%%%%%%%%%%%%%%%%%
\appendix
\section{Lorentz invariant formulation of the absorption cross-section}
\label{lorentz}

In the earlier version of our model \cite{LPPT1}, the relative yield $R$, 
the suppression factor $F$, and the absorption cross-section $\sigma_a$,
given by eqs.~(6), (7), and (3) of that paper, respectively, are
manifestly invariant under longitudinal boosts, since $\pt$ and $\sigma_0$
are boost invariant and the products $\rho_A\, dz_A$ and
$\rho_B\, dz_B$ are as well.
In the present formulation, the corresponding equations
\eqref{Eq:R}, \eqref{sig_a}, and \eqref{Eq:F} are valid in the CMS 
frame, but they can be also written in a Lorentz invariant way.

Prescription \eqref{sig_a} for the absorption cross-section
was formulated in the centre-of-mass
system. In this system, it says that $\sigma_a$ will be large if 
\begin{equation}
(t'-t)^2 \lesssim \frac{1}{\pt^2}\, .
\end{equation}
We treat only jets at central rapidity which do not move
longitudinally in CMS. The positions where the jet is created and
where it is hit by another nucleon are thus the same. Thus, the 
l.h.s.\ of the previous inequality can be written in a Lorentz invariant 
way: 
\begin{equation}
({x'}^\mu - x^\mu)(x^\prime_\mu - x_\mu) \lesssim \frac{1}{\pt^2} \, ,
\end{equation}
where ${x}^\mu$ and ${x'}^\mu$ are the four-vectors corresponding to events 
of jet creation and possible destruction, respectively. A Lorentz invariant
formula for the absorption cross-section for mid-rapidity jets
thus reads
\begin{equation}
\label{sig_inv}
\sigma_a(\pt,x^\mu,{x'}^\mu) = 
\sigma_0 \left ( \frac{1}{1 + \pt^2 ({x'}^\mu - x^\mu)(x^\prime_\mu - x_\mu)} 
\right )^2 \, .
\end{equation}

We can put this in a more suitable form. We neglect the movement
in transverse direction and write
\begin{equation}
({x'}^\mu - x^\mu)(x^\prime_\mu - x_\mu) = \Delta t^2 - \Delta z^2
\end{equation}
We denote:\\[0.3ex]
\begin{tabular}{rl}
 $y_j$ & the jet rapidity in a given frame,  \\[1ex]
 $y_A$ & the nucleon rapidity, same as nucleus rapidity, \\
$z^*$ & \parbox{9.8cm}{\flushleft 
the distance of the destroying nucleon from the place where 
the jet was produced, measured in the rest-frame of the nucleus.} 
\end{tabular}\\[2ex]
Then, the time between jet creation and its possible destruction 
can be written as
\begin{equation}
\Delta t = \frac{z^*}{\cosh y_A (\tanh y_A - \tanh y_j)} \, ,
\label{delt}
\end{equation}
and the distance between these two events is given by
\begin{equation}
\Delta z = \Delta t \, v_{\rm jet}  = 
\frac{z^* \, \tanh y_j}{\cosh y_A ( \tanh y_A - \tanh y_j)} \, .
\label{delz}
\end{equation}
Using these equations and after some algebra, we obtain
\begin{equation}
\label{Eq:t-z}
\Delta t^2 - \Delta z^2 = \frac{{z^*}^2}{\sinh^2(y_A - y_j)} \, .
\end{equation}
This form  depends only on difference of rapidities and is thus 
Lorentz invariant.

If we are only interested in midrapidity jets, we can replace
\[
y_j = \frac{y_A + y_B}{2}
\]
where $y_A$ and $y_B$ are rapidities of the colliding nuclei and
obtain the absorption cross section
\begin{equation}
\label{sim_a-rap}
\sigma_a = \sigma_0 \left ( 
\frac{1}{1+\frac{\pt^2\, {z^*}^2}{\sinh^2\left((y_A-y_B)/2\right )}} \right )^2
\, .
\end{equation}

%%%%%%%%%%%%%%%%%%%%%%%%%%%%%%%%%%%%%%%%%%%%%%%%%%%%%%%%%%%%%%%%%%
%%%%%%%%%%%%%%%%%%%%%%%%%%%%%%%%%%%%%%%%%%%%%%%%%%%%%%%%%%%%%%%%
%%%%%%%%%%%%%%%%%%%%%%%%%%%%%%%%%%%%%%%%%%%%%%%%%%%%%%%%%%%%%%%%

%%%%%%%%%%%%%%%%%%%%%%%%%%%%%%%%%%%%%%%%%%%%%%%%%%%%%%%%%%%%%%%%%%

\label{lastpage}

\end{document}